# An insight into spin-chain magnetism through Mössbauer spectroscopic investigations in Eu-doped $Ca_3Co_2O_6$ and $Ca_3CoRhO_6$

P.L. Paulose, Niharika Mohapatra and E.V. Sampathkumaran

*Tata Institute of Fundamental Research, Homi Bhabha Road, Colaba, Mumbai – 400005, India.*

We report the results of [151]Eu Mössbauer effect and magnetization measurements in the Eu-doped $Ca_3Co_2O_6$ and $Ca_3CoRhO_6$, which are of great current interest in the fields of spin-chain magnetism and geometrical frustration. We find that there is a pronounced increase in the line-width of the Mössbauer spectra below a certain temperature ($T^*$) which is well-above the one at which three-dimensional magnetic ordering features set in. This unusual broadening of the spectra indicates the existence of a characteristic temperature in these 'exotic' magnetic systems, attributable to the onset of *incipient* one-dimensional magnetic order. This is inferred from an intriguing correlation of $T^*$ with paramagnetic Curie temperature (a measure of intrachain coupling strength in these cases).

The investigation of magnetic behavior of one-dimensional (1D) spin-chain systems has been an active direction of research for the past several decades in condensed matter physics. It has been theoretically predicted long ago that a low dimensional (1D and 2D) chain with a non-zero exchange interaction can not order magnetically at finite temperatures [1]. While no exception for this rule could be found in bulk forms of the materials in the literature, a broad peak often appears in the plot of temperature (T) dependence of magnetic susceptibility ($\chi$) as a signature of short range intra-chain fluctuations [2]. It is however remarkable that a few years ago magnetic ordering in a 1D Co metal array has been realized at low temperatures as detected by X-ray magnetic circular dichroism [3]. It is therefore of great interest to readdress this fundamental issue in magnetism by searching for bulk materials in which strong signatures due to intra-chain magnetic ordering are seen.

In this article, we focus our attention on two spin-chain compounds, $Ca_3Co_2O_6$ and $Ca_3CoRhO_6$, belonging to $K_4CdCl_6$-type rhombohedral structure, which have attracted a lot of attention (including theoretical) in recent years due to a variety of novel features [4-11] exhibited by these compounds, a recent most fascinating finding being the onset of ferroelectricity at the magnetic transition [6]. In the former, NMR experiments by Sampathkumaran et al [4] conclusively establish that the Co at the (distorted) trigonal prismatic site is trivalent with high-spin $d^6$ coordination and the one at the (distorted) octahedral coordination is in a low-spin $d^6$ non-magnetic configuration. However, in the latter compound, trigonal-prismatically coordinated Co is in a divalent high-spin ($d^7$) configuration with Rh at the octahedral site in the low-spin $d^5$ magnetic configuration [7]. The ferromagnetic (Ising) spin-chains (made up of Co and Co-Rh respectively), isolated by Ca ions and running along c-axis, are arranged in a triangular fashion in the basal plane. See, for instance, figure 1 of Ref. 6, for a pictorial view of such crystallographic features. As a result of antiferromagnetic interchain interactions, the systems are topologically frustrated, leading to a 'not-so-common' 3D magnetic structure, called 'partially disordered antiferro (PDA) magnetic structure' below about $T_N$ (called $T_1$ hereafter) = 24 and 90 K respectively. There is another complex magnetic transition below about (called $T_2$) 10 and 30 K respectively. Thus, there are remarkable qualitative similarities in magnetic properties despite differences in the magnetism of the octahedral and trigonal prismatic sites in these two compounds. These materials attracted an attention with respect to application potential as well due to large thermopower [8]. It is notable that recent magnetization studies on the epitaxially grown thin films [9] and nanorods [10] of the former reveal additional magnetic transitions at somewhat higher temperatures (40 to 60 K). It becomes therefore important to address this issue of additional transition for bulk materials carefully for further theoretical advancement of knowledge of these 'exotic' compounds.

With these motivations, we have investigated $^{151}$Eu Mössbauer spectral behavior as a function of temperature in the Eu-doped samples, $Ca_{3-x}Eu_xCo_2O_6$ and $Ca_{3-x}Eu_xCoRhO_6$ ($x < 0.5$). In order to make sure that the doping does not cause any major perturbation of the magnetic properties, the focus is on the compositions with least amount of Eu, that is, $x = 0.1$. However, we present relevant data for one series, namely Rh-based series, for additional compositions to support a logical conclusion. The main point of observation is that, below a certain temperature, which is well above the long-range magnetic ordering temperature, there is a sudden onset of broadening of the spectra. This observation in the bulk form raises a question whether there exists a characteristic temperature ($T^*$), apparently pointing to *incipient* intra-chain magnetic ordering as judged by the existence of its correlation with paramagnetic Curie temperature ($\theta_p$).

The polycrystalline samples, $Ca_{2.9}Eu_{0.1}Co_2O_6$ and $Ca_{3-x}Eu_xCoRhO_6$ ($x=$ 0.1, 0.3 and 0.5), were prepared by solid state method as described in our earlier publications [4,5] starting from stoichiometric amounts of high purity (>99.99%) $CaCO_3$, $Co_3O_4$, $Eu_2O_3$ and Rh powder. The specimens obtained after a few heat treatments were found to be single phase by x-ray diffraction (Cu $K_\alpha$) within the detection limit (<1%) and the diffraction lines (in other words, the lattice constants) are found to smoothly shift with increasing Eu composition, exactly in the same way as noted for corresponding Y-substituted compositions [11]. In addition, the samples were confirmed to be homogeneous by scanning electron microscope. In order to magnetically characterize the samples, we have performed *dc* as well as *ac* magnetization (M) measurements (2-300 K) employing commercial magnetometers. $^{151}$Eu Mössbauer effect measurements were performed employing $^{151}$SmF$_3$ source (21.6 keV transition) at selected temperatures.

*We first present the results on $Ca_{2.9}Eu_{0.1}Co_2O_6$*. Since the features in *dc* and *ac* magnetization are found to be essentially the same as that of the parent compound reported in several places in the literature [4], we will not elaborate on these findings (see figure 1a). In brief, the magnetic susceptibility ($\chi$), for instance, measured with a magnetic field (H) of 5 kOe, exhibits a gradual increase with decrease of temperature down to 22 K, below which there is a sudden increase marking the onset of previously known PDA order for $x=$ 0.0 at 24 K. The onset of 3D-ordering has been is further supported by the presence of a peak in the heat-capacity data, shown in figure 1a. There is a bifurcation of the curves near 10 K for the zero-field-cooled (ZFC) and field-cooled (FC) conditions of the specimens, as demonstrated for H= 100 Oe in figure 1a. Isothermal M data (figure 1a, inset) measured at 15 K (that is, between $T_1$ and $T_2$) is non-hysteretic and show a step at about one-third of saturation magnetization typical of PDA structure. This step vanishes above $T_1$, as shown by the data at 30 K. At 1.8 K, M-H curve is hysteretic as expected [4]. AC $\chi$ exhibits (see figure 1b for the real part) an unusually large frequency dependence in the vicinity of the well-known magnetic transitions as in the parent compound. These results endorse that any knowledge obtained on this Eu-based composition is representative of the parent compound.

The Mössbauer spectra at few representative temperatures (300, 160, 100, 50, and 4.2 K) are shown in figure 2. The continuous lines through the data points are obtained by least squares fitting to a Lorentzian. The isomer shift is found to be about 1.0 mm/s falling in a range for trivalent Eu ions and it undergoes a usual weak thermal red shift with temperature. The line-width (W) at half-maximum at 300 K is 2.9 ± 0.1 mm/s and any possible quadrupolar contribution is accounted for within W. The most remarkable finding is that, while W remains essentially constant above about 100 K, there is a sudden increase as the temperature is lowered further with a pronounced effect below about 75 K (see figure 3). It is very surprising that this sets in a temperature range which is about thrice of (hither-to-known) magnetic-order onset temperature of $T_1$. However, interestingly, further increase in W below $T_1$ is sluggish as though the onset of partial antiferromagnetic ordering (2 out of 3) of spin-chains dampens the effect. For the aim of this article, we focus our further discussions on W behavior above $T_1$.

Generally speaking, W is expected to attain the paramagnetic value even in quasi 1D systems [12] as soon as three dimensional magnetic ordering disappears. We propose that the observed broadening arises from strong intra-chain exchange coupling, as the temperature around which the broadening appears prominently is nearly the same as that of $\theta_p$ of about 80 K inferred from the $\chi$ studies extended to temperatures well above 300 K (Maignan et al, Ref. 4). (From the positive sign of $\theta_p$, it is widely known that $\theta_p$ corresponds to intrachain ferromagnetic correlations in this family). With this in mind, we conclude that magnetic ordering within spin-chains tends to set in around 80 K. It is intriguing to note that it is static enough to be sensed [13] within the Mössbauer time scale of about $10^{-8}$ sec. As a result, there is a transferred hyperfine field ($H_{eff}$) to the Ca site from the neighbouring magnetically ordered Co ions. The broadening at 4.2 K corresponds to about 60 kOe. The corresponding value at 50 K is about 35 kOe, which is huge for this temperature range. Incidentally, the values of $H_{eff}$ at the Co site has been shown to be negligible or comparatively very small as demonstrated by doped-$^{57}$Fe Mössbauer effect measurements [14] or $^{59}$Co NMR measurements (Sampathkumaran et al, Ref. 4) respectively; this finding is sufficient to prove that doped-Eu does not go to chain sites, but to Ca site. Incidentally, within the conventional relaxation mechanism, ('dynamic picture'), it is rather difficult to understand the onset of abrupt broadening in a narrow temperature range and one would have expected a rather gradual (exponential) temperature dependence of W as demonstrated for doped-$^{57}$Fe in Ref.14. As a manifestation of incipent intra-chain ordering, the non-linear thermal excitation of moving domain walls (solitons or kinks) discussed in Ref. 15 for Ising spin-chain systems could result in the observed broadening, as demonstrated [16] for a similar line-broadening well above 3D-magnetic ordering temperature in the $^{57}$Fe Mössbauer spectra of $FeCl_2(pyridine)_2$. A change in the activation energy for the domain-wall propagation across can result in a corresponding change in the slope of the plot in figure 3 near $T_1$. In any case, Mössbauer data provide distinct evidence for the fact that one can define a $T^*$, related to incipient intra-chain magnetic ordering. Though correspondingly there is no well-defined anomaly in the $\chi(T)$ curve, one should take note of the observation that inverse $\chi(T)$ has been known [4] to show a deviation from high temperature linearity around this temperature possibly as a manifestation of incipient spin-chain ordering. It is also worth noting that there is a change in the functional form of temperature dependence of electrical resistivity around 80 K [17], the origin of which also could lie in this.

*We now turn to the results on $Ca_{3-x}Eu_xCoRhO_6$ to endorse the above conclusions.* As mentioned in the introduction, the magnetic behavior of $Ca_3Co_2O_6$ and $Ca_3CoRhO_6$ look similar (Hardy et al. Ref. 5), if one looks at the $\chi(T)$ normalized to respective $T_1$. However, it has been reported [5] that there is a distinct broad peak in $\chi(T)$ around 100 to 125 K in the latter, which has been attributed to intra-chain short-range correlations by traditional belief. This peak is more prominently seen for $x = 0.1$ with the downward shift of $T_1$ to 50K-range (from 90 K for $x = 0.0$) as indicated by a sudden upturn in $\chi$. This shift in $T_1$ is apparent from the fact that isothermal M (see inset of figure 4a) shows a weak step around 50 kOe at 35 K only, but not at 62 K [4]. Another magnetic transition in the vicinity of ($T_2 =$) 30 K is also visible (see figure 4a) in the dc $\chi(T)$ plot and weakly hysteretic M(H) behavior at 5 K is the same as that known for the parent compound.

Huge frequency dependence of ac $\chi$ [5,11], as in the parent compound, in the vicinity of long-range magnetic ordering are retained for this composition as well (see figure 4b). Therefore, qualitatively speaking, marginally doped Eu sample carries the physics of the parent compound. Therefore, we look at the Mössbauer spectra at selected temperatures and typical behavior of the spectra is shown in figure 5 (for 300, 160, 125, 100 and 4.2 K) for this composition. Focusing on the main point, W, plotted in figure 3, shows a pronounced increase below ($T^*=$) 125 K, which is far above $T_1$. Typical values of $H_{eff}$ estimated from W at 100 and 125 K are about 37 and 17 kOe respectively. In the absence of high temperature (>300 K) $\chi$ data, it is rather difficult to get a precise estimate of $\theta_p$ to bring out a correlation with W. Nevertheless, an inference based on the $\chi$ data in the range 250-300 K could be made. The sign of $\theta_p$ thus obtained is positive representing intra-chain coupling and the value obtained (from the range 250-300 K) is about 125 K. This value is marginally smaller than that (140 K, Ref. 11) for the parent compound and comparable to $T^*$. This correlation is further established by the following observation based on the Mössbauer data on other compositions (*x* = 0.3 and 0.5, the spectra are not shown here): The curve W(T) shifts to lower temperatures with increasing concentration of Eu (see figure 3), as though $T^*$ decreases and there is a corresponding decrease of $\theta_p$ (90 and 60 K) (nearly same trend as in Y substituted series, presented in Ref. 11). *A key support to the above interpretation is provided by the normalized plot, shown in the inset of figure 3:* That is, the plot of reduced linewidth, $(W-W_{300K})/(W_{4.2K}-W_{300K})$, versus normalized temperature, $T/T^*$, for all the samples studied is an universal curve. The values of $T^*$ thus are: 80 K for $Ca_{2.9}Eu_{0.1}Co_2O_6$; 150, 130 and 110 K for *x*= 0.1, 0.3 and 0.5 of $Ca_{3-x}Eu_xCoRhO_6$ respectively. These values follow the trends in $\theta_p$.

Finally, Loewenhaupt et al. [Ref. 5] reported the observation of a quasi-elastic line in the neutron diffraction data of $Ca_3CoRhO_6$ in the entire temperature range (not only above $T_1$, but also below $T_1$), arising from intra-chain interactions. In light of the present results, we have carefully reexamined the data presented in figure 5 of Ref. 5. An important point that has emerged is that the integrated intensity of the magnetic Bragg peak starts building actually at around 125 K – well above $T_1$ - (of course with a sharper increase below $T_1$ as expected) in the form of a 'knee' at the expense of the diffuse peak. The Diffuse Gaussian line also narrows below about 150 K. Thus there is already a signal for spin-chain magnetism at a temperature well above $T_1$ in the neutron diffraction as well, supporting present conclusions.

To conclude, we provide a microscopic experimental evidence for the existence of incipient magnetic order of spin-chains through (doped) $^{151}$Eu Mössbauer spectroscopic studies in the spin-chain systems, $Ca_3Co_2O_6$ and $Ca_3CoRhO_6$. The conclusion is based on an intriguing relationship between spectral broadening and the paramagnetic Curie temperature representing intra-chain coupling. In addition, the present results bring out the need to recognize a characteristic temperature in future theoretical formulation of these 'exotic' systems.

We thank Kartik K Iyer for his help while performing experiments.

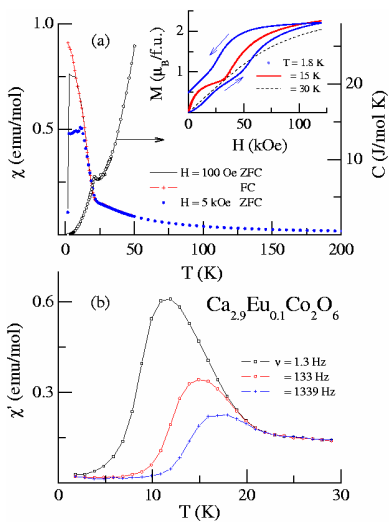

Figure 1:
 (a) *Dc* magnetic susceptibility (χ) as a function of temperature measured in a magnetic field of 5 kOe and of 100 Oe and (b) frequency (ν) dependence of real part *ac* χ, for $Ca_{2.9}Eu_{0.1}Co_2O_6$. The isothermal magnetization data are shown in the inset of (a). The heat-capacity data is also shown in (a).

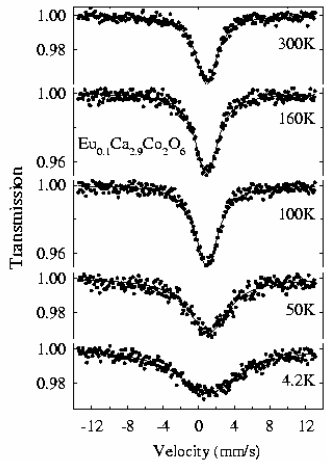

Figure 2:
 $^{151}$Eu Mössbauer spectra for $Ca_{2.9}Eu_{0.1}Co_2O_6$.

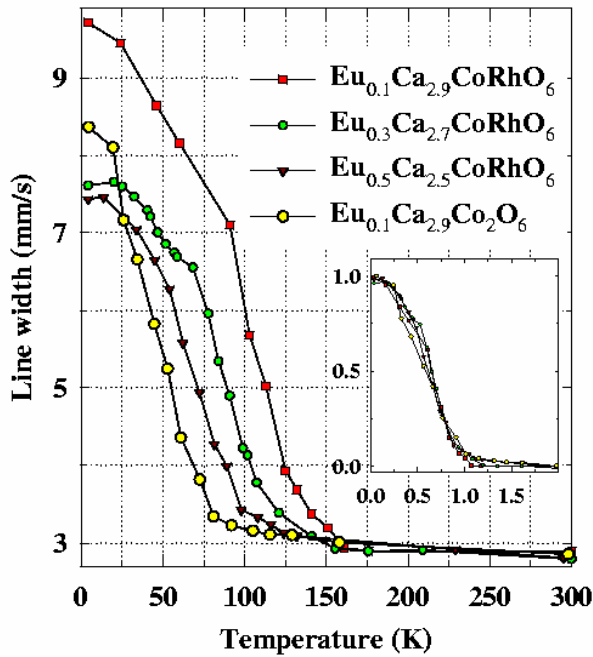

Figure 3: The line-width at half maximum (W) (± 0.1mm/s) of the Mössbauer spectra. The universal curve of the normalized data is shown by plotting (W-$W_{300K}$)/($W_{4.2K}$--$W_{300K}$) versus T/T$^*$ in the inset.

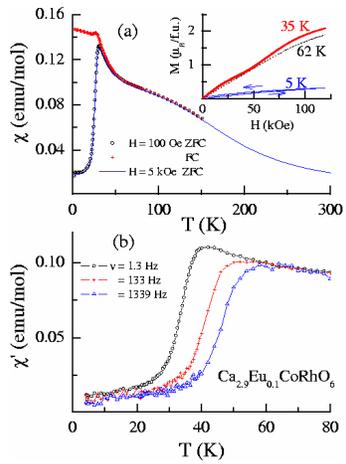

Figure 4:

The magnetization data (as in figure 1) for $Ca_{2.9}Eu_{0.1}CoRhO_6$. In *(a)*, the curves for H= 100 Oe and 5 kOe overlap.

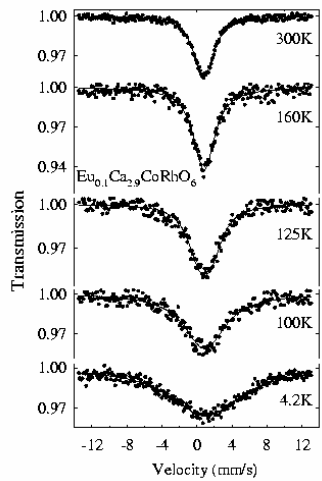

Figure 5:
$^{151}$Eu Mössbauer spectra for $Ca_{2.9}Eu_{0.1}CoRhO_6$.